\documentclass[submission, Proceedings]{SciPost}
\usepackage{tabularx}
\usepackage{graphicx}
\usepackage{multirow}
\usepackage{pifont}
\newcommand{\cmark}{\ding{51}}
\begin{document}

\begin{center}{\Large \textbf{
Monte Carlo, fitting and Machine Learning for Tau leptons
}}\end{center}

\begin{center}
V. Cherepanov\textsuperscript{1},
E. Richter-Was\textsuperscript{2},
Z. Was\textsuperscript{3}\textsuperscript{$^\dagger$}
\end{center}

\begin{center}
{\bf 1} Institut Pluridisciplinaire Hubert Curien (IPHC), 67037  Strasbourg, France
\\
{\bf 2} Institute of Physics, Jagellonian University, 30-348 Krakow, Lojasiewicza 11, Poland
\\
{\bf 3} Institute of Nuclear Physics, Polish Academy of Sciences,  Radzikowskiego 152, PL-31342 Krakow, Poland
\\
  {\bf $^\dagger$} Speaker, email address zbigniew.was{\it at}ifj.edu.pl\\
\end{center}

\begin{center}
  IFJPAN-IV-2018-18,\; \;\;\;\today
\end{center}

\definecolor{palegray}{gray}{0.95}
\begin{center}
\colorbox{palegray}{
  \begin{tabular}{rr}
  \begin{minipage}{0.05\textwidth}
    \includegraphics[width=8mm]{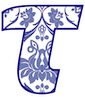}
  \end{minipage}
  &
  \begin{minipage}{0.82\textwidth}
    \begin{center}
    {\it Proceedings for the 15th International Workshop on Tau Lepton Physics,}\\
    {\it Amsterdam, The Netherlands, 24-28 September 2018} \\
    \href{https://scipost.org/SciPostPhysProc.1}{\small \sf scipost.org/SciPostPhysProc.Tau2018}\\
    \end{center}
  \end{minipage}
\end{tabular}
}
\end{center}


\section*{Abstract}
{\bf
Status of    
$\tau$ lepton decay Monte Carlo generator {\tt TAUOLA}, and its main recent
applications are reviewed.  It is underlined, that in 
recent efforts on development of new hadronic currents, the multi-dimensional 
nature of distributions of  the experimental data must be taken with a great care. 
Studies for $H \to \tau\tau ; \tau \to hadrons$ indeed demonstrate
 that multi-dimensional nature of distributions is
 important and available  for evaluation of observables where $\tau$ leptons
 are used to constrain 
experimental data.
For that part of the presentation,
use of the {\tt TAUOLA} program for phenomenology of $H$ and $Z$ decays at LHC 
is discussed, in particular in the context of the Higgs boson parity measurements
with the use of Machine Learning  techniques.
Some additions, relevant  for QED lepton pair emission and electroweak corrections are mentioned as well. 
}

\vspace{10pt}
\noindent\rule{\textwidth}{1pt}
\tableofcontents\thispagestyle{fancy}
\noindent\rule{\textwidth}{1pt}
\vspace{10pt}

\section{Introduction}
\label{intro}

It is  thirty years since the first versions of {\tt TAUOLA} package
\cite{Jadach:1990mz,Jezabek:1991qp,Jadach:1993hs,Golonka:2003xt} for simulation
of $\tau$-lepton decays and
{\tt PHOTOS} \cite{Barberio:1990ms,Barberio:1994qi,Golonka:2005pn} for simulation of QED radiative corrections
in decays became public. The interfaces of these programs are designed in a way to fulfil needs of different groups of users.
The bulk of the code remain written and maintained by
the same (main) authors, but contributions originating from other researchers became integrated over the time too.
These programs  became parts of  a wide range of applications, sometimes consisting of quite 
complicated simulation chains. Some versions of these codes at the moment require an independent maintenance.
Some of those code  modifications are not prepared and maintained by the program authors. This is often
the case of
{\tt TAUOLA}.  From the design point of view new versions differ little, usually in numerical values of some
constants or analytical form of hadronic currents. Nonetheless, these  changes  are  of a great value because they
often depend on fits to experimental data,
sometimes even not fully made public by the experiments. Variants may archive
details of some other 
$\tau$-lepton phenomenological projects, these topics were presented in previous $\tau$ conferences.

Our presentation is organized as follows:
Section 2  is devoted to the discussion of recent changes in
 {\tt TAUOLA} package. 
This point was already discussed in  \cite{Was:2014zma} and  in \cite{Chrzaszcz:2016fte}, that is why, we will be brief.
In Section 3 we  summarize recent developments for   {\tt PHOTOS} Monte Carlo of
radiative corrections in decays. 
Section 4 is devoted to  applications of {\tt TAUOLA}
  for hard processes with final state $\tau$ leptons. In particular, for
construction of spin observables and evaluation of their sensitivity.
 The modern techniques like 
Machine Learning (ML),   were found to be important in analyses of HEP data.
 In this context we discuss  {\tt TauSpinner} algorithm, which was found to be useful for evaluation 
of observable for Higgs boson parity measurement. We mention 
 other applications or tests for this tool; in particular in the domain of 
algorithm of calculating spin states of $\tau$ pairs in events where
high $p_T$ jets are present in $pp$ collisions. The following two sections provide
examples for applications in the domain of precision measurements and Higgs boson CP parity
evaluation at LHC.
 Summary Section 7 closes the presentation.

\section{Currents and structure of  {\tt TAUOLA} Monte Carlo}

The {\tt TAUOLA} and {\tt PHOTOS} represent long term projects, that is why no substantial
changes were introduced 
since the 
$\tau$ conferences of 2014 and 2016 \cite{Was:2014zma,Was:2016goo}. Nowadays, the {\tt C++ }
became dominant in  many segments of  the code for programs and for the tests as well. 
The structure defining  part of the code was modified, to prepare for  more modular code
organization. In fact only small changes in the text of the code were necessary, that is why,
numerical tests if introduced changes were bug free were simple. Now, once decision to
re-write 
the whole code into  {\tt C++} is finally taken, it can be performed   in quick and well
controlled steps, each to be tested separately. See Ref.~\cite{Chrzaszcz:2016fte} for
more explanations. Let us point, that language change for {\tt TAUOLA} is more complex than it was
for  {\tt PHOTOS} \cite{Davidson:2010ew}.
This is because of the three phenomenology aspects of the work which need to be
taken into account:
\begin{itemize}
\item  Development of physic assumptions and later of the code, for new versions of
  hadronic currents of all or some $\tau$ decay 
  channels.
\item Preparation  of 
    experimental data for fits. Question of background control or dimensionality of the
    distributions is of a great importance. For the optimal performance participation of
    physicists involved in experiments data analysis is important.
\item Preparations of algorithms 
      and choice  of distributions to be used in fits. We will return to this point later in the
      presentation.
\end{itemize}
Synchronization of work on these tasks and  evolution of the code need to be assured.
That is not easy, as different
researchers are involved in each of these activities.

In the following sections,  Optimal Variables and ML techniques are evaluated
for purposes of Higgs and Z phenomenology. We expect  it  of importance for
modeling  of $\tau$ decays as well.
Especially in this case, proper evaluation of multidimensional distributions
used for fits and in particular systematic uncertainties may be of a great
importance. We study if  Monte Carlo
matrix element dependent event weights may be useful.

Finally let us recall some limitations  on physics precision of the {\tt TAUOLA}
resulting from the program design.
Up to a precision level of about 0.2\% 
choice of hadronic currents   play the
central role for systematic uncertainties. That is why, corresponding parts of the program
represent  well defined, easy to replace, building 
block.
Confrontation of model's predictions with experimental data with the help 
of multi-dimensional distribution has to be central for the future project developments.
This was pointed already in Ref.~\cite{Asner:1999kj} and its importance is clear from our  
experience 
of work on hadronic currents for $\tau \to 3\pi\nu$ decay modes as well~\cite{Was:2014zma,Was:2016goo}.

\section{{\tt PHOTOS} Monte Carlo for bremsstrahlung: \newline its systematic uncertainties}
\def\CCol{{\tt SANC}}
Over the last two years no major upgrades for functionalities 
were introduced into {\tt PHOTOS} Monte Carlo, except introduction 
of emission of lepton pairs. Published documentation of the program~\cite{Davidson:2010ew}, correspond to the present day head  version\cite{HeadPhotos}.

Numerical tests for pair emission algorithm are  published \cite{Antropov:2017bed}, 
and updates on the program necessary due to evolution how event records
contents have to be understood are continuously updated, see 
{\tt changelog.txt} of Ref.~\cite{HeadPhotos}. A possibility for the user
modification of the $Z\to l^+l^-\gamma$ matrix element is prepared.
The appropriate matrix element can be replaced by the user own.
In this way, e.g. effects of non-standard-model matrix elements
or leading contributions of the loop electroweak corrections, can be studied.

\section{{\tt TAUOLA } - hard process - {\tt TauSpinner} algorithm }
The packages such as {\tt TAUOLA} or {\tt PHOTOS} are rarely
used alone.  Also  tests of the programs can not be performed fully
independently of users' projects. Thus  stand-alone tests are not sufficient.
Typically, user applications rely on other libraries of programs  as well,
which in size may
surpass largely {\tt TAUOLA} or {\tt PHOTOS} themselves. The complete
simulation chains
consist not only of segments for so called `truth' i.e.
physics processes based on theoretical predictions,
but emulate detector response as well. 
Physics interest is usually not focused on properties of  simulation
chains,
but on  intermediate state properties such as mass,
coupling constants or parity.
If intermediate state properties can be modified by the experimental user
and consequences of such changes for detector responses as well,
work for defining sensitive observables can be  simplified.

For the final states with $\tau$ leptons, methods to manipulate their
spin state  can be used to optimize the measurements. The $\tau$ is the only
lepton of its spin accessible to the measurements.
From the first paper \cite{Czyczula:2012ny} {\tt TauSpinner} was oriented
for such applications.
The response   due to changes in
 $Z, W$ or $H$ decays, represents a valuable information.
Let us review the status. General structure of the
tool was recently documented in~\cite{Przedzinski:2018ett}, but
let us nonetheless point to the general idea.
The program is calculating {\it weights} corresponding to changes of the physics
assumption. Ratios of matrix elements squared for the compared
options is used for that purpose.
As an input,
events stored on the data file are used. No changes for kinematical
configurations are introduced. In that respect
nothing  has changed since 2016 $\tau$ conference and remain
as in Ref.~\cite{Was:2016goo}, however new applications were
documented in Refs.~\cite{Kalinowski:2016qcd,Bahmani:2017wbm}. With these publications,
possibility to use matrix elements, for parton level processes
with the two outgoing jets accompanying $\tau$ pair, was introduced. Another
new option was 
to introduce, with the help of {\tt TauSpinner}, weights that account for the electroweak loop corrections
to Drell-Yan processes~\cite{Richter-Was:2018lld}. 

With the help of  {\tt TauSpinner} we could evaluate observable to study Higgs
boson parity,
in its cascade decay with intermediate $\tau$ leptons \cite{Jozefowicz:2016kvz}. More precisely;
we have investigated the potential for measuring the CP state of
the Higgs boson in $H \to \tau \tau$ decay, 
with consecutive  $\tau$-lepton decays in the channels: $\tau^{\pm} \to \rho^{\pm} \nu_{\tau}$ 
and $\tau^{\pm} \to a_{1}^{\pm} \nu_{\tau}$ combined. Subsequent decays 
$\rho^{\pm} \to \pi^{\pm} \pi^{0}$,  $ a_{1}^{\pm} \to \rho^{0} \pi^{\pm}$ and
 $ \rho^{0}\to \pi^{+}\pi^{-}$ were taken into account. 

We have extended 
method of Ref.~\cite{Bower:2002zx} first.
Also in the case of the cascade decays of $\tau \to a_1 \nu$, 
information on the CP state of Higgs boson can be extracted  
from the acoplanarity angles. However for
$ a_{1}^{\pm} \to \rho^{0} \pi^{\pm},\; \rho^{0} \to \pi^+ \pi^-$  cascade,
up to four planes 
can be defined,  thus  16 distinct acoplanarity angle distributions
 are available for
 $H \to \tau \tau \to a_1^{+} a_1^{-} \nu \nu$.
 Each of such distributions carry some information but 
 it is laborious  to evaluate overall sensitivity using previous approach.

We have investigated the sensitivity with the help of
Machine Learning (ML) techniques. 
Since the 2016 $\tau$ conference, we have in part taken into account
ambiguities resulting from the detector 
uncertainties and  background contamination, see Ref.~\cite{Barberio:2017ngd}.
Sensitivity largely survived.

Usefulness of ML methods and $\tau \to 3\pi \nu$ decays for Higgs boson parity 
measurement can be understood as an example. It required analysis of
multi-dimensional distributions simultaneously for signatures and background
processes. The case of Higgs boson parity
remained physics-wise simple, the signatures to be distinguished, were
described by analytically clear and simple weights. The examples
which we describe below, of interest in themselves, may also
lead to helpful solutions for physics of $\tau$ decays.
There, of course theoretical description is challenging, but thanks
to better control of interferences (thanks to multidimensional distributions)
new insight may become available. 
Let us  now return to the examples of the high energy domain.
Here, the ML  and   more classical (but leading to intuition) solutions
based on so called Optimal Variables techniques were used.

\section{Towards Optimal Variables for $Z/H \rightarrow \tau\tau$ spin observables}

The knowledge of the $\tau$ lepton kinematic is essentially important for the spin analysis. An angle between the $\tau$ lepton
flight direction and a neutrino in its decay is  a powerful spin analyzer. This angle can be reconstructed unambiguously only if
both the total momentum and the flight direction of the $\tau$ lepton are reconstructed.  Unlike in $e^{+}e^{-}$ collisions in pp collisions
at LHC there is no beam energy constraint, however it is possible to place a reasonably good estimate on the neutrino momenta 
in the decays $Z/H \rightarrow \tau\tau$ \cite{Bianchini:2014vza,Elagin:2010aw,Cherepanov:2018npf}.
In this section two examples of the {\tt TauSpinner} application to the analysis of the longitudinal and transverse spin of $\tau$ leptons in the decays $Z/H \rightarrow \tau\tau$ assuming that the 
whole kinematic of the decay is available are discussed.

\subsection{Longitudinal $\tau$ polarization in the decay $Z \rightarrow \tau\tau$}
Let us start with the example of longitudinal $\tau$  polarization,
the observable important for the precision
tests of the Standard Model.
The difference of the neutral weak couplings to the right- and left-handed fermions results in the polarization of the fermion-antifermion pairs produced in
the decay of the Z bosons. A measurement of the $\tau$ leptons polarization  at LHC provides an independent and complementary determination
of the effective weak mixing angle $\sin^{2}\theta^{\tau}_{eff}$ as well as a test of the lepton universality of the weak neutral current.  
The measurement of the $\tau$ polarization requires a knowledge of the $\tau$ spin state, this can be concluded by analyzing the angular distributions
of the $\tau$ decay products with respect to the $\tau$ flight direction. 
Following the notations of \cite{Jadach:1993hs} the partial width of any $\tau$ decay is given by:

\begin{equation}
d\Gamma = \frac{1}{2M}|\bar{M}|^{2} (1+ h_{\mu}s^{\mu})dLips,
\end{equation}
where M is the $\tau$ mass, $|\bar{M}|^{2}$ spin averaged matrix element, $s$ -  four-vector of the $\tau$ polarization and   polarimetric vector $h$ is
a function of the $\tau$ momentum and the decay products. The standard abbreviation 
for Lorentz invariant phase-space integration element $dLips$ is used.
And angle $\cos\theta_h$ (further referred as $\omega_h$) between the polarimetric vector $h$ and the $\tau$ flight direction, as seen from the 
$\tau$ rest frame, carries the full information about the spin of the $\tau$ (assuming only longitudinal spin component of the $\tau$ leptons).

\begin{figure}[h]
  \begin{center}
    \includegraphics[width=0.45\textwidth]{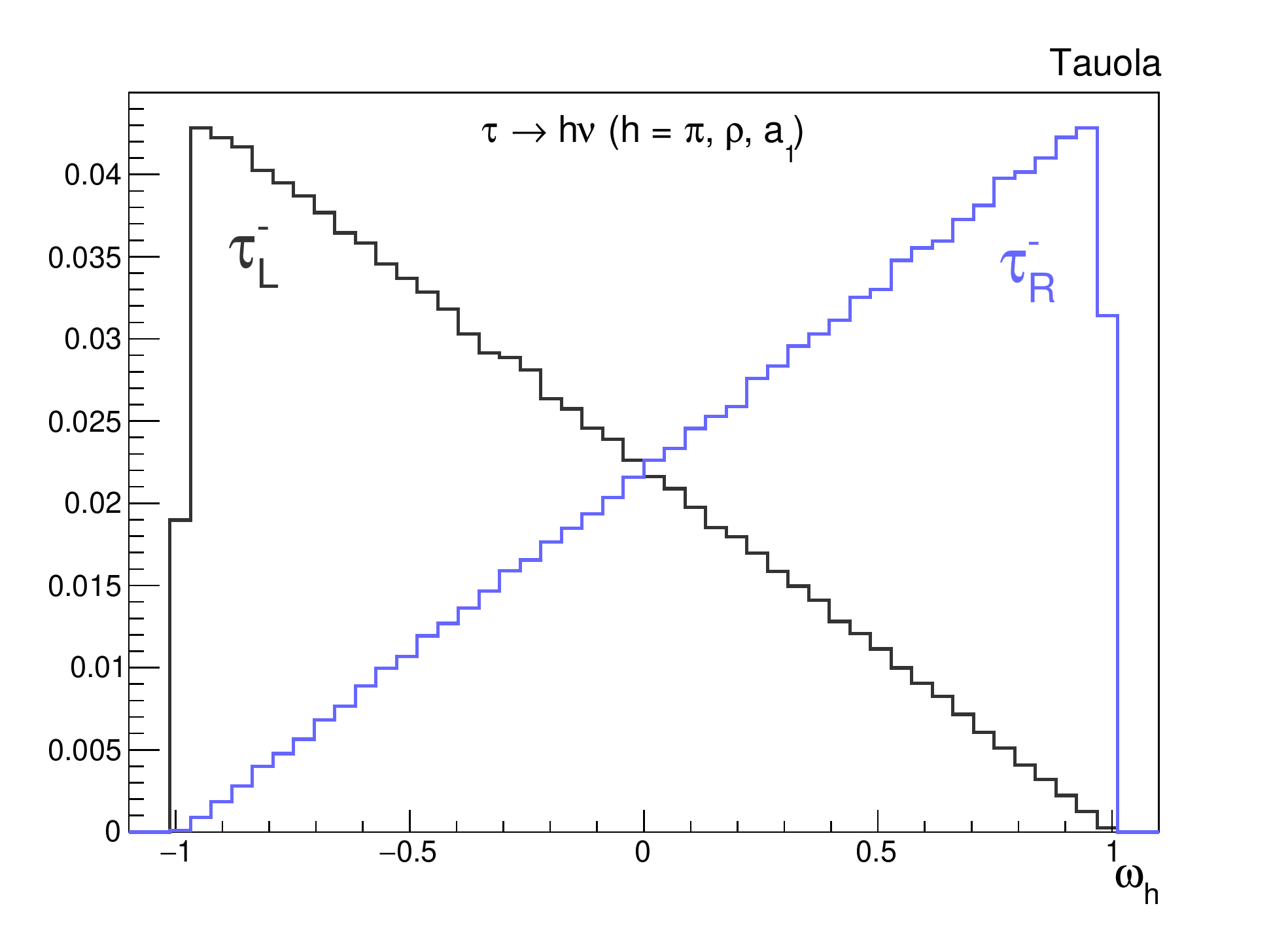}  
    \includegraphics[width=0.45\textwidth]{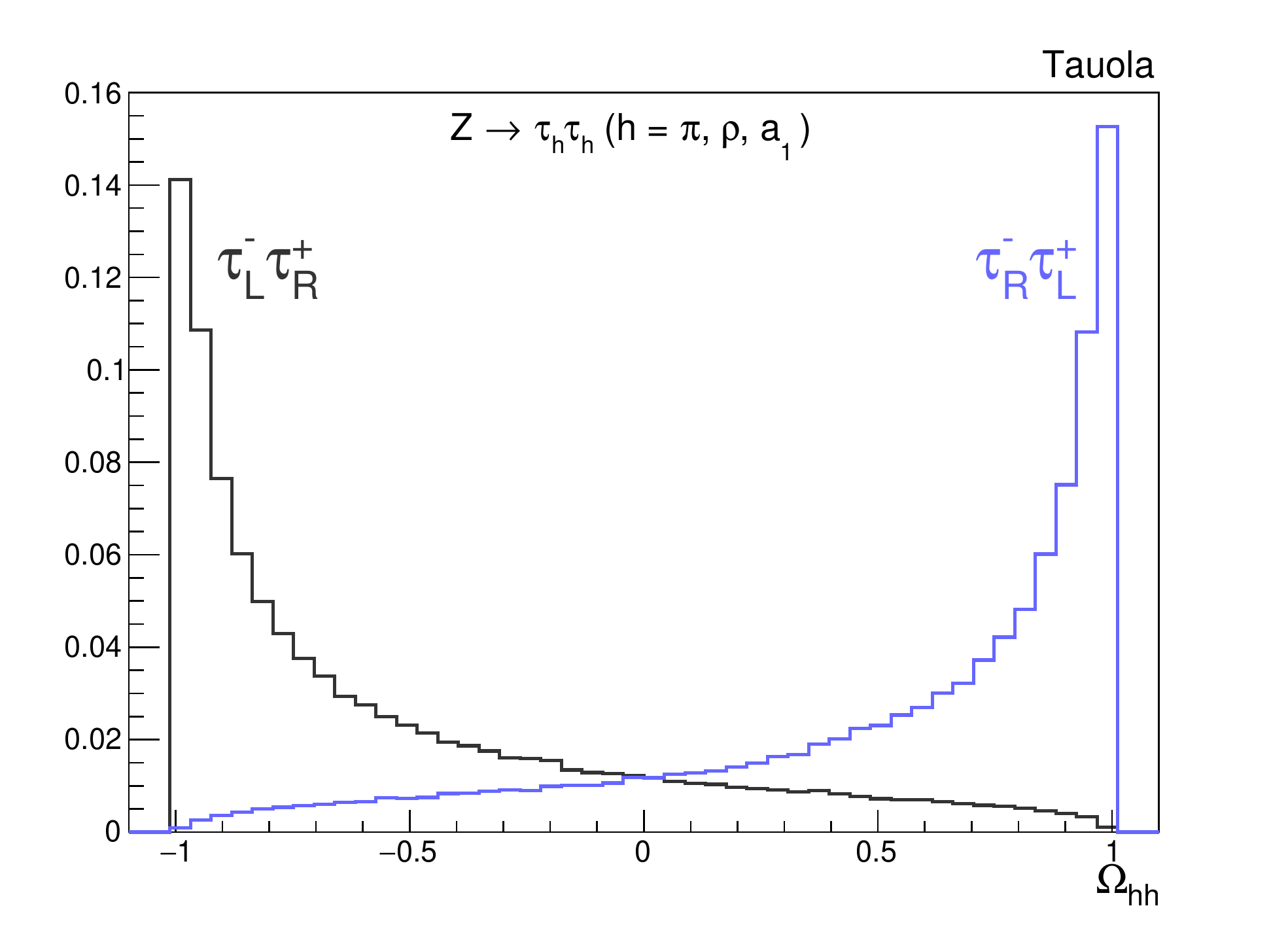}  
    \caption{The distribution of $\omega_{h}$ ({\it Left}) and $\Omega_{hh}$  ({\it Right}) for negative (black)  and positive (cyan) $\tau^{-}$ helicity. The distribution is similar for all considered $\tau$ decay channel.}
      \label{fig:omegas}
  \end{center}
\end{figure}

An explicit form of the polarimetric vector for $\tau$ decays,  can be well identified by CMS \cite{Chatrchyan:2008aa} or ATLAS \cite{Aad:2008zzm} detectors 
($\tau^{-} \rightarrow \pi^{-}\nu$, $\pi^{-}\pi^{0}\nu$ and $\pi^{-}\pi^{-}\pi^{+}\nu$),
it can be found in \cite{Jadach:1993hs}. An analyzing power of the $\tau$ 
polarization can be further gained noting that the helicity states of both $\tau$ leptons in the decay $Z \rightarrow \tau\tau$ are almost 100\% anti-correlated. Denoting 
$\omega^{1}_{h}$ and $\omega^{2}_{h}$  to be the observables for both $\tau$ leptons a combined observable is given by \cite{Davier:1992nw}:

\begin{equation}
\Omega_{hh} = \frac{\omega^{1}_{h} + \omega^{2}_{h}}{1+\omega^{1}_{h}\omega^{2}_{h}}.
\end{equation}

The distribution of a single $\omega_h$ and $\Omega_{hh}$ for the decays $\tau^{-} \rightarrow \pi^{-}\nu$, $\pi^{-}\pi^{0}\nu$ and $\pi^{-}\pi^{-}\pi^{+}\nu$  is shown in Fig.~\ref{fig:omegas}.
It should be noted that  only in the decay $\tau \rightarrow \pi^{-}\pi^{-}\pi^{+}\nu$ there is a model dependence that comes from the imperfect knowledge of the $a^{-}_{1} \rightarrow \pi^{-}\pi^{-}\pi^{+}$ decay structure. It has been shown that assuming the CLEO parametrization of the $a^{-}_{1} \rightarrow \pi^{-}\pi^{-}\pi^{+}$ decay \cite{Browder:1999fr} the modeling systematic uncertainty on the the $\tau$ polarization measurement is negligibly small and will not limit the precision \cite{Cherepanov:2018wop}.

\subsection{Transverse spin correlation in the decay $H \rightarrow \tau\tau$ and a quest for an Optimal Observable of the Higgs boson CP parity}

There is no longitudinal net polarization of $\tau$ leptons in the decay $H \rightarrow \tau\tau$, however  the transverse spin
correlation of $\tau$ leptons in the decay $H \rightarrow \tau\tau$ might reveal the information on the CP structure of this decay. 

\begin{figure}[h]
  \begin{center}
    \includegraphics[width=0.45\textwidth]{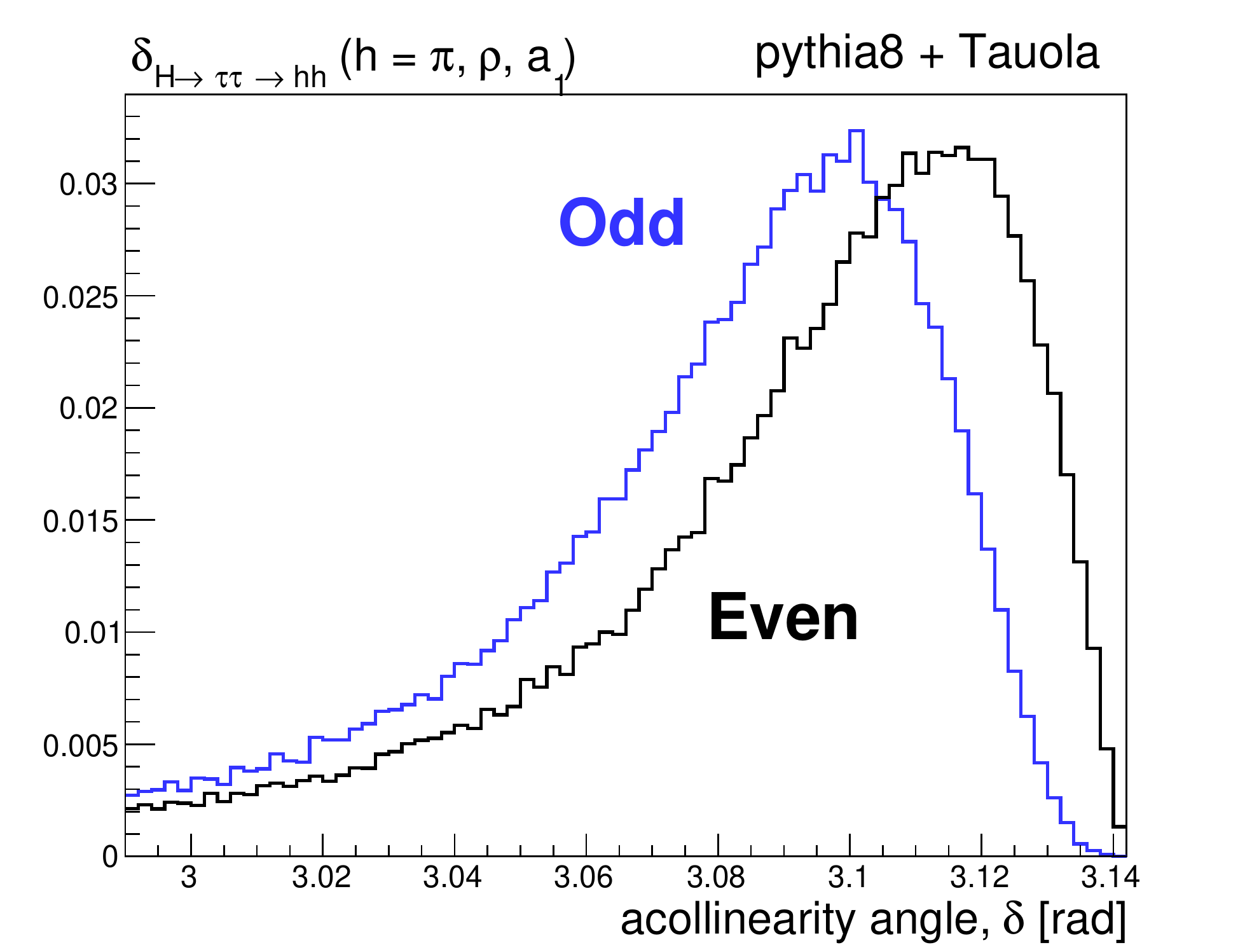}  
    \includegraphics[width=0.45\textwidth]{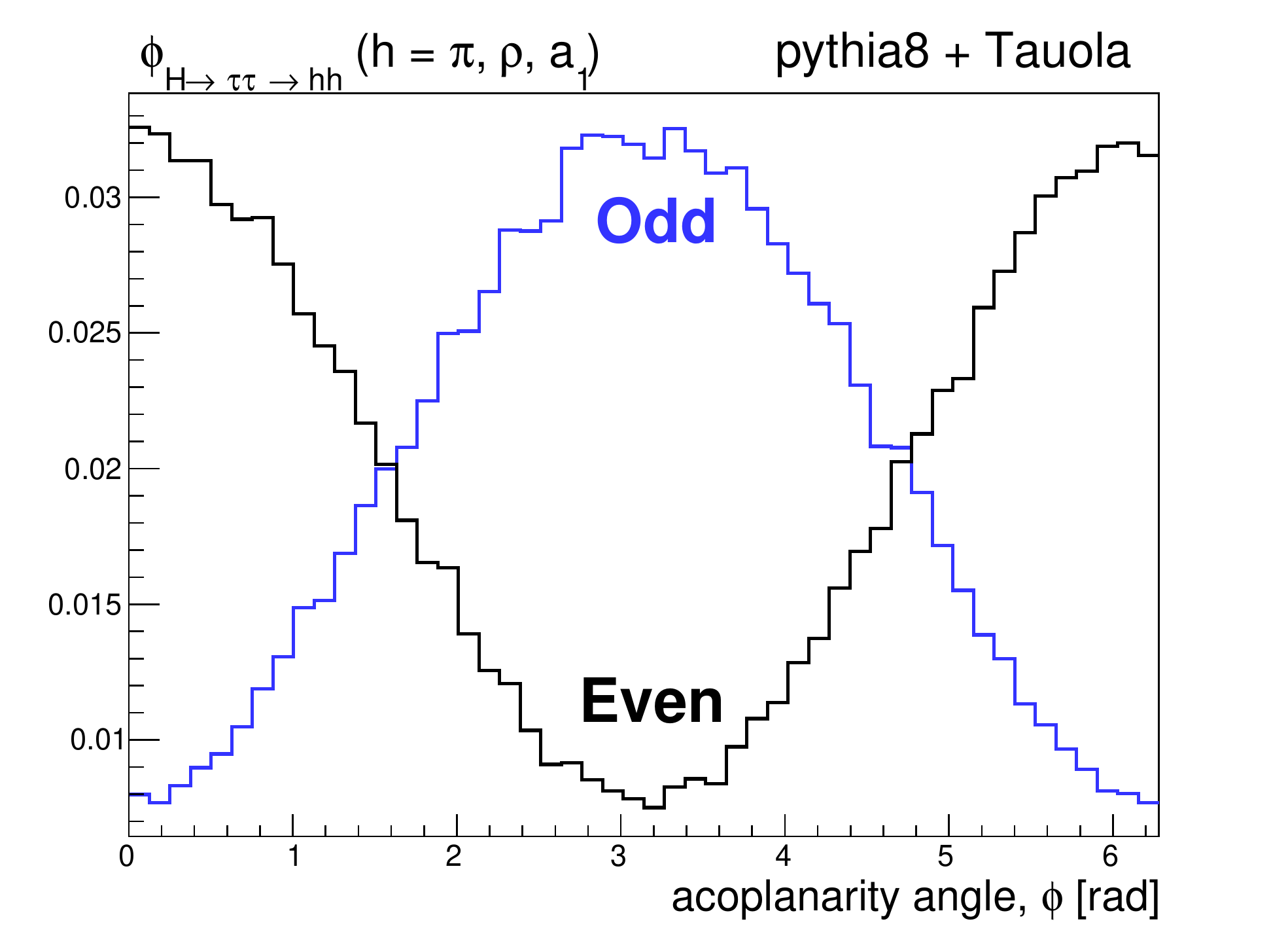}  
    \caption{The distribution of $\delta$ ({\it Left}) and $\phi$ ({\it Right}) for CP even and odd amplitude. The distributions are identical for considered $\tau \rightarrow \pi\nu$, $\tau \rightarrow \rho\nu$, $\tau \rightarrow a_{1}\nu$ decays. }
      \label{fig:hcp}
  \end{center}
\end{figure}

Similarly to the longitudinal spin analysis the analyzing power of the transverse spin correlation can be maximized by considering the full kinematic of the $\tau$ pair. Denoting $\vec{h}_{1}$ and $\vec{h}_{2}$ as the polarimetric vectors of both $\tau$ leptons in their rest frame of $\tau$ pair:

\begin{eqnarray}\label{eq:acolacompl}
\delta &=& \arccos(\vec{h}_{1}\cdot\vec{h}_{2}), \nonumber\\
\phi^{\ast} &=& \arccos(\vec{k}_{1}\cdot\vec{k}_{2}),
\end{eqnarray}
where $\vec{k}_{1,2}= \frac{\vec{h}_{1,2}\times\vec{n}_{1,2}}{|\vec{h}_{1,2}\times\vec{n}_{1,2}|}$. The $\vec{n}_{1,2}$ are the unit vectors pointing along the direction of the $\tau^{-}$ and $\tau^{+}$ in the $H$ rest frame. The angle $\phi^{\ast}$ varies in the range from 0 to $\pi$ only, the signed acoplanarity  angle $\phi$ ($0 \leq \phi \leq 2\pi$) can be defined as\footnote[1]{In the printed version of the proceeding the definitions for the  angles $\delta$ and $\phi^{\ast}$  are incomplete/misleading; fixed June 28, 2019. }:

\begin{equation}
\phi =
\begin{cases}
    \phi^{\ast}                    & \text{if } (\vec{h}_{1}\times \vec{h}_{2})\cdot\vec{n}_{1} \leq  0\\
    2\pi -\phi^{\ast}              & \text{if } (\vec{h}_{1}\times \vec{h}_{2})\cdot\vec{n}_{1} >  0
\end{cases}
\end{equation}

 The distributions of $\delta$ and $\phi$ for even and  odd Higgs boson CP state are shown in Fig.~\ref{fig:hcp}. They are are uniform for all considered $\tau$
decays and carry the full analyzing power. However, the analyzing power in the experiment will be naturally diluted by the detector effects and limited resolution. The best performance can be expected when both $\tau$ leptons decay into three charged pions. Three charged tracks in the detector offer robust
reconstruction of the point of each $\tau$ decay (no flight length of the quickly decaying $a_{1}$ resonance) and the high multiplicity of pp collisions allows to determine the point of interaction. This information, imposed as an additional constraint, can significantly improve the performance of algorithms for kinematic reconstruction of the $\tau$ leptons momenta.

In this approach all experimental complexity is hidden in the measurement of
$\vec{h}_{1}$ and $\vec{h}_{2}$. The  $\delta$ or $\phi$ take the role of
Optimal Variables. That is very helpful for physics intuition.

\section{From Optimal Variables to Machine Learning approach}
Let us quote an example of numerical result taken from Ref.~\cite{Jozefowicz:2016kvz} and Ref.~\cite{Barberio:2017ngd}. For the
analysis of the sensitivity of $H \to \tau\tau ; \; \tau \to \nu_\tau 2(3)\pi$
we have used ML technique. This path was explored, because form the previous
studies \cite{Bower:2002zx} we could expect that the signature will depend
of the distribution features embedded over multidimensional space. As a part of
our approach was to evaluate reliability of the new for us techniques, we
performed study of prototypes for the possible elements of construction of
Optimal Variables, defining hyperspace  of this multidimensional space.
We recall results,  the Table \ref{tab:DeepLearn} for the case when detector
smearings were not taken into account, and  Table~\ref{table:baseline} where
they were taken into account. The encouraging results  have been
obtained.

Let us now recall a few details, that may be useful if an attempt to apply
similar techniques will be directed, e.g. toward evaluation of models and fits
of the $\tau$ decay matrix elements.
The discussion of the systematic uncertainties of the results
could be simplified, because for each event we could calculate matrix elements
for the confronted assumption. Complexity of the resulting weights was
the results of $\tau$ decays, but the part describing vertex $H \to \tau \tau$
was rather simple and of clear character.

In case of $\tau$ decays complexity and ambiguities in definition of matrix
elements as a function of details of the intermediate energy strong
interaction models will be inevitably larger. However, even in this case
corresponding to the confronted assumption weights will be available.
Already now, this is technically prepared and embedded in {\tt TauSpinner}
algorithm. Note that different assumption of $\tau$ decay models
may lead to relatively similar predictions for one dimensional distribution,
but largely differ for the multidimensional ones, see for example ~\cite{Was:2015laa}.

In our applications for the Higgs boson we have confronted two assumptions
for the Higgs boson  CP parity states, in most cases it was choice between
scalar and pseudo-scalar variants only. However, the forthcoming solutions
based on ML approach will have the possibility of evaluating best
performing parameters of some models too. This will open the way
for fits of $\tau$ decays themselves.
We leave such solutions for the forthcoming works.

\begin{table*}
\resizebox{\textwidth}{!}{
  \begin{tabular}{lrrrr}
\hline
\hline
  Features/variables      & Decay mode: $\rho^{\pm}- \rho^{\mp}$    &  Decay mode: $a_1^{\pm} - \rho^{\mp}$   & Decay mode:  $a_1^{\pm} - a_1^{\mp}$  \\ 
                          & $\rho^{\pm} \to \pi^{0}\ \pi^{\pm}$   &  $ a_{1}^{\pm} \to \rho^{0} \pi^{\mp},\ \rho^{0} \to  \pi^{+} \pi^{-}$  
                                                               &  $ a_{1}^{\pm} \to \rho^{0} \pi^{\pm},\ \rho^{0} \to  \pi^{+} \pi^{-}$             \\ 
                          &                                   &  $\rho^{\mp} \to \pi^{0}\ \pi^{\mp}$   &                                         \\
  \hline
  True classification                             & 0.782       &  0.782          &  0.782    \\
  $\varphi^*_{i,k}$                                       & 0.500       &  0.500         &  0.500     \\
  $\varphi^*_{i,k}$ and $y_i, y_k$                             & 0.624       &  0.569          &  0.536    \\
    4-vectors                                                & 0.638       &  0.590          &  0.557    \\
  $\varphi^*_{i,k}$, 4-vectors                            & 0.638       &  0.594          &  0.573    \\
  $\varphi^*_{i,k}$, $y_i, y_k$ and $m^2_i, m^2_k$                   & 0.626       &  0.578          &  0.548    \\
  $\varphi_{i,k}^*$, $y_i$, $y_k$, $m^2_i$, $m^2_k$ and 4-vectors         & 0.639       &  0.596          &  0.573    \\
\hline
\hline
\end{tabular}
}
\caption{ Average probability $p_i$ 
(calculated as explained in  Ref.~\cite{Jozefowicz:2016kvz})
that a model predicts correctly event $x_i$ to be of a  type $A$ (scalar),
with training being performed for separation between type $A$ and $B$ (pseudo-scalar). Different sets of variables were used as an input. The $\varphi^*_{i,k}$,
$y_i, y_k$ can be understood as approaches to construct expert variables, but
as we could observe, their use did not brought improvements with respect to
the case when 4-vectors of the $\tau$ decay products were used alone, provided
proper reference frame was used. 
Results taken from Ref.~\cite{Jozefowicz:2016kvz}.
\label{tab:DeepLearn}}
\end{table*}
{\scriptsize
\begin{table*}
    \centering
        \begin{tabular}{cccccccc}
        \multicolumn{4}{c}{\small Features}  & \multicolumn{1}{c}{\multirow{2}{20 mm}{ Ideal $\pm$ (stat)}} & \multirow{2}{35 mm}{Smeared $\pm$ (stat) $\pm$ (syst)} & \multicolumn{1}{c}{\multirow{2}{20 mm}{}} \\ \cline{1-4}
        $\phi^*$ & 4-vec &  $y_i$ & $m_i$ & \multicolumn{1}{c}{}    & \multicolumn{1}{c}{}             &                      \\
        \hline
        \multicolumn{7}{c}{$a_1-\rho$ Decays}\\
        \cmark & \cmark & \cmark & \cmark & $0.6035 \pm 0.0005$ & $0.5923 \pm 0.0005 \pm 0.0002$ &  \\
        \cmark & \cmark & \cmark & -      & $0.5965 \pm 0.0005$ & $0.5889 \pm 0.0005 \pm 0.0002$ &  \\
        \cmark & \cmark & -      & \cmark & $0.6037 \pm 0.0005$ & $0.5933 \pm 0.0005 \pm 0.0003$ & \\
        -      &{\bf \cmark} & -      & -      & $0.5971 \pm 0.0005$ & $0.5892 \pm 0.0005 \pm 0.0002$ &  \\
        \cmark & \cmark & -      & -      & $0.5971 \pm 0.0005$ & $0.5893 \pm 0.0005 \pm 0.0002$ &  \\
        \cmark & -      & \cmark & \cmark & $0.5927 \pm 0.0005$ & $0.5847 \pm 0.0005 \pm 0.0002$ &  \\
        \cmark & -      & \cmark & -      & $0.5819 \pm 0.0005$ & $0.5746 \pm 0.0005 \pm 0.0002$ &  \\
        \multicolumn{7}{c}{$a_1-a_1$ Decays}\\
        \cmark & \cmark & \cmark & \cmark & $0.5669 \pm 0.0004$ & $0.5657 \pm 0.0004 \pm 0.0001$ &  \\
        \cmark & \cmark & \cmark & -      & $0.5596 \pm 0.0004$ & $0.5599 \pm 0.0004 \pm 0.0001$ & \\
        \cmark & \cmark & -      & \cmark & $0.5677 \pm 0.0004$ & $0.5661 \pm 0.0004 \pm 0.0001$ &  \\
        -      &{\bf  \cmark} & -      & -      & $0.5654 \pm 0.0004$ & $0.5641 \pm 0.0004 \pm 0.0001$ &  \\
        \cmark & \cmark & -      & -      & $0.5623 \pm 0.0004$ & $0.5615 \pm 0.0004 \pm 0.0001$ & \\
        \cmark & -      & \cmark & \cmark & $0.5469 \pm 0.0004$ & $0.5466 \pm 0.0004 \pm 0.0001$ & \\
        \cmark & -      & \cmark & -      & $0.5369 \pm 0.0004$ & $0.5374 \pm 0.0004 \pm 0.0001$ & \\
        \end{tabular}
        \caption{Area Under the Curve is shown for the Neural Networks (NN) studied to
separate  scalar and pseudo-scalar hypotheses.
          Inputs with a \cmark are used. 
    Results in column  ``Ideal" - from NNs trained/used with particle-level simulation,
    in column  ``Smeared" - from NNs trained/used with smearing.
    NN trained on smeared samples when  used on exact samples give similar results as
    ``Ideal" what in this case mean no detector smearing.
    Result taken from Ref.~\cite{Barberio:2017ngd}. Reference frame as for Table~\ref{tab:DeepLearn}.
}
    \label{table:baseline}
\end{table*}
}

\section{Summary and future possibilities}

There was not much of the development for the implementation of $\tau$ lepton decays
embedded in {\tt TAUOLA} library. The new version of the program implementation
was archived and finally published \cite{Chrzaszcz:2016fte}.
The status of
associated projects: {\tt TAUOLA universal interface } and {\tt TauSpinner}
was reviewed. Also new results for 
the high-precision version of  {\tt PHOTOS} for QED radiative corrections in
decays, were presented, in particular algorithm for emission of additional light lepton
pairs was supported with documented tests.

Some details of presentation of the {\tt TAUOLA} general-purpose {\tt C++} interface
was given and its applications were shown. Use of weighted events was demonstrated
to be useful for studies of LHC phenomenology in domain of $\tau$ lepton polarization
observables for precision tests of Standard Model as well as for Higgs boson CP parity.
Complementary methods of Optimal Variables as well as of ML approaches,
both useful for phenomenological applications were presented and example results were shown.

\vskip 2 mm
\centerline{\bf Acknowledgments}

{\small 
The work on {\tt TAUOLA} would not be possible without continuous help an encouragements
from experimental colleagues.
The work of all  co-authors of the papers devoted to  {\tt TAUOLA} development 
was of great importance.
I hope, that it is clearly visible from  my contribution. 
This project was supported in part from funds of Polish National
Science Centre under decisions DEC-2017/27/B/ST2/01391 and by
Institut National de Physique Nucl\'eaire et de Physique des Particules / CNRS.
}




\begin{thebibliography}{10}
\providecommand{\url}[1]{\texttt{#1}}
\providecommand{\urlprefix}{URL }
\expandafter\ifx\csname urlstyle\endcsname\relax
  \providecommand{\doi}[1]{doi:\discretionary{}{}{}#1}\else
  \providecommand{\doi}{doi:\discretionary{}{}{}\begingroup
  \urlstyle{rm}\Url}\fi
\providecommand{\eprint}[2][]{\url{#2}}

\bibitem{Jadach:1990mz}
S.~Jadach, J.~H. K\"{uhn} and Z.~W\c{a}s,
\newblock \emph{Tauola: A library of monte carlo programs to simulate decays of
  polarized tau leptons},
\newblock Comput. Phys. Commun. \textbf{64}, 275 (1990).

\bibitem{Jezabek:1991qp}
M.~Je\.zabek, Z.~W\c{a}s, S.~Jadach and J.~H. K\"{u}hn,
\newblock \emph{The tau decay library tauola, update with exact o(alpha) qed
  corrections in tau $\to$ mu. (e) neutrino anti-neutrino decay modes},
\newblock Comput. Phys. Commun. \textbf{70}, 69 (1992).

\bibitem{Jadach:1993hs}
S.~Jadach, Z.~Was, R.~Decker and J.~H. K\"{u}hn,
\newblock \emph{The tau decay library tauola: Version 2.4},
\newblock Comput. Phys. Commun. \textbf{76}, 361 (1993).

\bibitem{Golonka:2003xt}
P.~Golonka \emph{et~al.},
\newblock \emph{{The tauola-photos-F environment for the TAUOLA and PHOTOS
  packages, release II}},
\newblock Comput. Phys. Commun. \textbf{174}, 818 (2006),
\newblock \doi{10.1016/j.cpc.2005.12.018},
\newblock \eprint{hep-ph/0312240}.

\bibitem{Barberio:1990ms}
E.~Barberio, B.~van Eijk and Z.~W\c{a}s,
\newblock \emph{Photos: A universal monte carlo for qed radiative corrections
  in decays},
\newblock Comput. Phys. Commun. \textbf{66}, 115 (1991).

\bibitem{Barberio:1994qi}
E.~Barberio and Z.~W\c{a}s,
\newblock \emph{Photos: A universal monte carlo for qed radiative corrections.
  version 2.0},
\newblock Comput. Phys. Commun. \textbf{79}, 291 (1994).

\bibitem{Golonka:2005pn}
P.~Golonka and Z.~Was,
\newblock \emph{{PHOTOS Monte Carlo: A precision tool for QED corrections in Z
  and W decays}},
\newblock Eur. Phys. J. \textbf{C45}, 97 (2006),
\newblock \eprint{hep-ph/0506026}.

\bibitem{Was:2014zma}
Z.~Was,
\newblock \emph{{The $\tau$ leptons theory and experimental data: Monte Carlo,
  fits, software and systematic errors}},
\newblock Nucl. Part. Phys. Proc. \textbf{260}, 47 (2015),
\newblock \doi{10.1016/j.nuclphysbps.2015.02.010},
\newblock \eprint{1412.2937}.

\bibitem{Chrzaszcz:2016fte}
M.~Chrzaszcz, T.~Przedzinski, Z.~Was and J.~Zaremba,
\newblock \emph{{TAUOLA of $\tau$ lepton decays-framework for hadronic
  currents, matrix elements and anomalous decays}},
\newblock Comput. Phys. Commun. \textbf{232}, 220 (2018),
\newblock \doi{10.1016/j.cpc.2018.05.017},
\newblock \eprint{1609.04617}.

\bibitem{Was:2016goo}
Z.~Was,
\newblock \emph{{Tau lepton production and decays: perspective of
  multi-dimensional distributions and Monte Carlo methods}},
\newblock Nucl. Part. Phys. Proc. \textbf{287-288}, 15 (2017),
\newblock \doi{10.1016/j.nuclphysbps.2017.03.035},
\newblock \eprint{1611.08838}.

\bibitem{Davidson:2010ew}
N.~Davidson, T.~Przedzinski and Z.~Was,
\newblock \emph{{PHOTOS Interface in C++: Technical and Physics
  Documentation}},
\newblock Comput. Phys. Commun. \textbf{199}, 86 (2016),
\newblock \doi{10.1016/j.cpc.2015.09.013},
\newblock \eprint{1011.0937}.

\bibitem{Asner:1999kj}
D.~Asner \emph{et~al.},
\newblock \emph{{Hadronic structure in the decay $\tau^- \to \tau_\nu \pi^-
  \pi^0 \pi^0$ and the sign of the $\tau_\nu$ helicity}},
\newblock Phys.Rev. \textbf{D61}, 012002 (2000),
\newblock \doi{10.1103/PhysRevD.61.012002},
\newblock \eprint{hep-ex/9902022}.

\bibitem{HeadPhotos}
{\tt Photos++} website http://photospp.web.cern.ch/photospp/.

\bibitem{Antropov:2017bed}
S.~Antropov, A.~Arbuzov, R.~Sadykov and Z.~Was,
\newblock \emph{{Extra lepton pair emission corrections to Drell-Yan processes
  in PHOTOS and SANC}},
\newblock Acta Phys. Polon. \textbf{B48}, 1469 (2017),
\newblock \doi{10.5506/APhysPolB.48.1469},
\newblock \eprint{1706.05571}.

\bibitem{Czyczula:2012ny}
Z.~Czyczula, T.~Przedzinski and Z.~Was,
\newblock \emph{{TauSpinner Program for Studies on Spin Effect in tau
  Production at the LHC}},
\newblock Eur.Phys.J. \textbf{C72}, 1988 (2012),
\newblock \doi{10.1140/epjc/s10052-012-1988-z},
\newblock \eprint{1201.0117}.

\bibitem{Przedzinski:2018ett}
T.~Przedzinski, E.~Richter-Was and Z.~Was,
\newblock \emph{{Documentation of TauSpinner algorithms -- program for
  simulating spin effects in tau-lepton production at LHC}}  (2018),
\newblock \eprint{1802.05459}.

\bibitem{Kalinowski:2016qcd}
J.~Kalinowski, W.~Kotlarski, E.~Richter-Was and Z.~Was,
\newblock \emph{{Production of $\tau $ lepton pairs with high $p_T$ jets at the
  LHC and the TauSpinner reweighting algorithm}},
\newblock Eur. Phys. J. \textbf{C76}(10), 540 (2016),
\newblock \doi{10.1140/epjc/s10052-016-4361-9},
\newblock \eprint{1604.00964}.

\bibitem{Bahmani:2017wbm}
M.~Bahmani, J.~Kalinowski, W.~Kotlarski, E.~Richter-Was and Z.~Was,
\newblock \emph{{Production of $\tau \tau jj$ final states at the LHC and the
  TauSpinner algorithm: the spin-2 case}},
\newblock Eur. Phys. J. \textbf{C78}(1), 10 (2018),
\newblock \doi{10.1140/epjc/s10052-017-5480-7},
\newblock \eprint{1708.03671}.

\bibitem{Richter-Was:2018lld}
E.~Richter-Was and Z.~Was,
\newblock \emph{{The TauSpinner approach for electroweak corrections in LHC $Z
  \to \ell\ell$ observables}}  (2018),
\newblock \eprint{1808.08616}.

\bibitem{Jozefowicz:2016kvz}
R.~Jozefowicz, E.~Richter-Was and Z.~Was,
\newblock \emph{{Potential for optimizing the Higgs boson CP measurement in H
  $\to \tau \tau$ decays at the LHC including machine learning techniques}},
\newblock Phys. Rev. \textbf{D94}(9), 093001 (2016),
\newblock \doi{10.1103/PhysRevD.94.093001},
\newblock \eprint{1608.02609}.

\bibitem{Bower:2002zx}
G.~Bower, T.~Pierzchala, Z.~Was and M.~Worek,
\newblock \emph{{Measuring the Higgs boson's parity using tau $\to$ rho nu}},
\newblock Phys.Lett. \textbf{B543}, 227 (2002),
\newblock \doi{10.1016/S0370-2693(02)02445-0},
\newblock \eprint{hep-ph/0204292}.

\bibitem{Barberio:2017ngd}
E.~Barberio, B.~Le, E.~Richter-Was, Z.~Was, D.~Zanzi and J.~Zaremba,
\newblock \emph{{Deep learning approach to the Higgs boson CP measurement in $H
  \to \tau \tau$ decay and associated systematics}},
\newblock Phys. Rev. \textbf{D96}(7), 073002 (2017),
\newblock \doi{10.1103/PhysRevD.96.073002},
\newblock \eprint{1706.07983}.

\bibitem{Bianchini:2014vza}
L.~Bianchini, J.~Conway, E.~K. Friis and C.~Veelken,
\newblock \emph{{Reconstruction of the Higgs mass in $H \to \tau\tau$ Events by
  Dynamical Likelihood techniques}},
\newblock J. Phys. Conf. Ser. \textbf{513}, 022035 (2014),
\newblock \doi{10.1088/1742-6596/513/2/022035}.

\bibitem{Elagin:2010aw}
A.~Elagin, P.~Murat, A.~Pranko and A.~Safonov,
\newblock \emph{{A New Mass Reconstruction Technique for Resonances Decaying to
  di-tau}},
\newblock Nucl.Instrum.Meth. \textbf{A654}, 481 (2011),
\newblock \doi{10.1016/j.nima.2011.07.009},
\newblock \eprint{1012.4686}.

\bibitem{Cherepanov:2018npf}
V.~Cherepanov and A.~Zotz,
\newblock \emph{{Kinematic reconstruction of $Z/H \rightarrow \tau\tau$ decay
  in proton-proton collisions }}  (2018),
\newblock \eprint{1805.06988}.

\bibitem{Chatrchyan:2008aa}
S.~Chatrchyan \emph{et~al.},
\newblock \emph{{The CMS Experiment at the CERN LHC}},
\newblock JINST \textbf{3}, S08004 (2008),
\newblock \doi{10.1088/1748-0221/3/08/S08004}.

\bibitem{Aad:2008zzm}
G.~Aad \emph{et~al.},
\newblock \emph{{The ATLAS Experiment at the CERN Large Hadron Collider}},
\newblock JINST \textbf{3}, S08003 (2008),
\newblock \doi{10.1088/1748-0221/3/08/S08003}.

\bibitem{Davier:1992nw}
M.~Davier, L.~Duflot, F.~Le~Diberder and A.~Rouge,
\newblock \emph{{The Optimal method for the measurement of tau polarization}},
\newblock Phys.Lett. \textbf{B306}, 411 (1993),
\newblock \doi{10.1016/0370-2693(93)90101-M}.

\bibitem{Browder:1999fr}
T.~E. Browder \emph{et~al.},
\newblock \emph{{Structure functions in the decay tau-+ ---> pi-+ pi0 pi0
  neutrino(tau)}},
\newblock Phys. Rev. \textbf{D61}, 052004 (2000),
\newblock \doi{10.1103/PhysRevD.61.052004},
\newblock \eprint{hep-ex/9908030}.

\bibitem{Cherepanov:2018wop}
V.~Cherepanov and W.~Lohmann,
\newblock \emph{{Methods for a measurement of $\tau$ polarization asymmetry in
  the decay $Z\rightarrow \tau\tau$ at LHC and determination of the effective
  weak mixing angle}}  (2018),
\newblock \eprint{1805.10552}.

\bibitem{Was:2015laa}
Z.~Was and J.~Zaremba,
\newblock \emph{{Study of variants for Monte Carlo generators of $\tau
  \rightarrow 3\pi \nu $ decays}},
\newblock Eur. Phys. J. \textbf{C75}(11), 566 (2015),
\newblock \doi{10.1140/epjc/s10052-016-4295-2, 10.1140/epjc/s10052-015-3780-3},
\newblock [Erratum: Eur. Phys. J.C76,no.8,465(2016)],
\newblock \eprint{1508.06424}.

\end{thebibliography}

\nolinenumbers

\end{document}